Epitaxial Co-Deposition Growth of CaGe$_2$ Films by Molecular Beam Epitaxy for Large Area Germanane

Igor V. Pinchuk,[1,2] Patrick M. Odenthal,[2] Adam S. Ahmed,[1] Walid Amamou,[2] Joshua E. Goldberger,[3] and Roland K. Kawakami[1,2,*]

[1]Department of Physics, The Ohio State University, Columbus, OH 43210

[2]Department of Physics and Astronomy, University of California, Riverside, CA 92521

[3]Department of Chemistry, The Ohio State University, Columbus, OH 43210

*email: kawakami.15@osu.edu

**ABSTRACT**

Two-dimensional crystals are an important class of materials for novel physics, chemistry, and engineering. Germanane (GeH), the germanium-based analogue of graphane (CH), is of particular interest due to its direct band gap and spin-orbit coupling. Here, we report the successful co-deposition growth of CaGe$_2$ films on Ge(111) substrates by molecular beam epitaxy (MBE) and their subsequent conversion to germanane by immersion in hydrochloric acid. We find that the growth of CaGe$_2$ occurs within an adsorption-limited growth regime, which ensures stoichiometry of the film. We utilize *in situ* reflection high energy electron diffraction (RHEED) to explore the growth temperature window and find the best RHEED patterns at 750 °C. Finally, the CaGe$_2$ films are immersed in hydrochloric acid to convert the films to germanane. Auger electron spectroscopy of the resulting film indicates the removal of Ca and RHEED patterns indicate a single-crystal film with in-plane orientation dictated by the underlying Ge(111) substrate. X-ray diffraction and Raman spectroscopy indicate that the resulting films are indeed germanane. *Ex situ* atomic force microscopy (AFM) shows that the grain size of the germanane is on the order of a few micrometers, being primarily limited by terraces induced by the miscut of the Ge substrate. Thus, optimization of the substrate could lead to the long-term goal of large area germanane films.

## I. INTRODUCTION

The discovery and mechanical exfoliation of single layer graphene has led to a revolution in nanoscale materials science by enabling a plethora of new physics and chemistry in two dimensions (2D) that is not possible in conventional three-dimensional materials.[1-3] Beyond graphene, there is great interest in exploring 2D materials that offer properties not found in graphene such as a native band gap, strong spin-orbit coupling, and strong exciton confinement.[4] Of particular interest is germanane (GeH), the germanium-based analogue of graphane (CH), which has recently been synthesized as a bulk single-crystal and exfoliated onto insulating substrates.[5] This was achieved by first synthesizing bulk crystals of layered Zintl-phase $CaGe_2$, then submerging the crystals in hydrochloric acid (HCl) to de-intercalate the Ca atoms (Fig. 1). This results in stacked layers of 2D germanane sheets, where each sheet consists of Ge atoms arranged in a buckled honeycomb lattice and $sp^3$-hybridized due to covalent bonds to hydrogen as shown in Fig 1(a).

Germanane has several properties that make it extremely attractive for science and engineering. First, it has a direct band gap of ~1.5 eV and is theoretically capable of high mobility transport (~18,000 $cm^2$/Vs), which make it favorable for optoelectronics and nanoscale digital electronics.[5] Second, it is thermodynamically stable at room temperature and is resistant to oxidation, which is advantageous for exploratory physics studies as well as future device applications. Finally, the larger spin-orbit coupling associated with heavier element Ge (as compared to C in graphene) opens up new opportunities to explore spintronics in 2D such as manipulation of spin by internal spin-orbit fields, spin Hall effect, and quantum spin Hall effect.[6-8]

An important challenge is the growth of large area, single-crystalline germanane films. The most successful effort to date has been the synthesis of $CaGe_2$ films by depositing Ca films on Ge(111) and then performing an *in situ* post-anneal.[9, 10] The $CaGe_2$ films were then converted to germanane via HCl de-intercalation as described above.[5, 11] While this provides an important baseline, a much stronger approach would be to perform molecular beam epitaxy (MBE) growth of $CaGe_2$ films by co-depositing Ca and Ge atoms. This would provide much greater control over the film thickness as compared to the post-anneal method and may lead to higher quality films. Further, the MBE co-deposition approach could lead to vertically-stacked heterostructures of germanane, silicane (SiH), and their alloys through the growth of $CaGe_2$-$CaSi_2$ heterostructures and subsequent HCl de-intercalation. However, to our knowledge, the co-deposition growth of $CaGe_2$ or $CaSi_2$ has not yet been demonstrated.

In this paper, we report the successful co-deposition growth of $CaGe_2$ films on Ge(111) substrates by MBE. We find that the surface structure of co-deposited $CaGe_2$ films is better than films prepared by post-annealing, based on a comparison of *in situ* reflection high energy electron diffraction (RHEED) patterns. Furthermore, we investigate the effect of growth temperature and find the best RHEED patterns at 750 °C. Finally, the $CaGe_2$ films are immersed in HCl to de-intercalate the Ca atoms, and characterization by x-ray diffraction and Raman spectroscopy show that the resulting films are indeed germanane. *Ex situ* atomic force microscopy (AFM) indicates that the grain size is on the order of a few micrometers, being limited primarily by terraces induced by the Ge substrate. With optimization of the substrate, it may be possible to realize the long-term goal of large area germanane. These results signal a major advance for 2D semiconductors and layered van der Waals heterostructures for novel electronic, optical, and spintronic devices.

## II. EXPERIMENTAL DETAILS

MBE growth is performed in an ultrahigh vacuum (UHV) chamber with base pressure of $2 \times 10^{-10}$ torr. Elemental Ge and Ca are evaporated from thermal effusion cells with high purity Ge (99.9999%, Alfa Aesar) and Ca (99.99%, Sigma Aldrich) source materials. Growth rates are determined by a quartz deposition monitor. All films are grown on undoped Ge(111) single-sided polished wafers with an average wafer thickness of 0.35 mm and an orientation tolerance of 0.5 degrees (University Wafer). The starting 2" diameter wafer is cleaved into smaller pieces which then undergo chemical etching to remove surface oxides/metals and replace them with a thin protective oxide film. Etching involves a sequence of steps beginning with submersion into a 10:1 mixture of $H_2O$:$NH_4OH$ for 60 sec followed by 60 sec in a 10:1 mixture of $H_2O$:$H_2SO_4$. Finally, substrates are put into a $H_2O_2$ solution for 60 sec before being rinsed by de-ionized (DI) water and inserted into the UHV chamber. The substrate is annealed at 650 °C for 30 min as measured by a thermocouple located near the substrate. *In situ* RHEED is used to monitor the sample surface throughout the growth and annealing process. The RHEED beam voltage is 15 kV unless otherwise noted. Annealing thermally desorbs the protective oxide from the Ge(111) surface. With the substrate still at 650 °C, a 5 nm Ge buffer layer is deposited with typical rates of ~2 Å/min as measured by a quartz microbalance deposition monitor. This procedure is followed for all $CaGe_2$ films grown in this study.

### III. RESULTS AND DISCUSSION

The co-deposition growth of $CaGe_2$ films on Ge(111) is illustrated by the RHEED patterns in Figure 2. Figure 2(a) shows RHEED patterns of a Ge(111) surface with beam along the $[11\bar{2}0]$ and $[1\bar{1}00]$ in-plane directions (left and right images, respectively), after thermal annealing to remove the surface oxide. Figure 2(b) shows respective patterns after the deposition of the 5 nm Ge buffer layer. The important feature here is sharpening of the RHEED diffraction streaks, indicating a flat and ordered Ge surface. In addition, the appearance of 1/4 and 1/8 order diffraction peaks in the $[11\bar{2}0]$ direction indicate the well-established Ge 2×8 reconstruction of a high quality, clean surface.[12-14]

The strategy for co-deposition growth of $CaGe_2$ is to try to achieve the adsorption-limited growth regime, similar to the MBE growth of GaAs, EuO, $Bi_2Se_3$, and other materials.[15-19] In such growths, one of the elements is supplied at overpressure and any excess will re-evaporate from the surface (volatile species) so that the growth rate is determined by deposition of the other element (stable species). If such a growth window exists, then good stoichiometry is obtained. For example, in GaAs growth the As (volatile species) is supplied at overpressure and the growth rate of GaAs is determined by the deposition of Ga (stable species). For the case of $CaGe_2$, the evaporation temperature of Ge is ~1100 °C and of Ca is ~500 °C (based on typical effusion cell temperatures). Therefore, Ca may act as the volatile, re-evaporating species. Furthermore, the bulk phase diagram for Ca-Ge alloys indicates that the $CaGe_2$ phase is stable for temperatures below ~800 °C.[20] Thus, there is the possibility for an adsorption-limited growth window between 500 °C and 800 °C.

For the adsorption-limited growth, the sample temperature is raised to 750 °C and the Ca and Ge fluxes are $8.1\times10^{12}$ atoms/cm$^2$s and $1.2\times10^{13}$ atoms/cm$^2$s, respectively. The flux ratio of Ca:Ge is 0.68 which has a higher Ca flux than the stoichiometric ratio of 0.5 for CaGe$_2$ to provide the required excess Ca for adsorption-limited growth. Here, we operationally define the growth rate of the CaGe$_2$ film as the rate determined by the Ge flux. We note that this is a nominal rate because it does not take into account the possible additional thickness of CaGe$_2$ due to interdiffusion and reaction of Ca into the Ge(111) substrate.[9,10] In the remainder of the paper, we cite the nominal film thickness associated with the deposited Ge. The Ge shutter is opened first, with the calcium shutter opening 5 sec later. Approximately 10 sec after the opening of the Ca shutter, the RHEED pattern shows a sequence of rapid changes in the surface reconstruction until the stable patterns shown in Figure 2(c) are achieved after another 20 sec. This pattern exhibits a 1×3 reconstruction, which remains for the duration of the CaGe$_2$ co-deposition (for thicknesses at least as high as 150 nm). As we verify later, this is the RHEED pattern for CaGe$_2$. Some important features need to be highlighted here. As the Ge(111) pattern changes into the CaGe$_2$ pattern, the integer order diffraction peaks remain strongly visible (in both directions). This shows that the main structure of the surface remains unchanged, as expected for the transition from Ge(111) to the CaGe$_2$ structure. As Ca is introduced into the bulk Ge lattice, the Ge(111) planes maintain their overall structure but every two Ge planes are separated by a Ca layer. We can visualize the process by examining Figures 1(b) and (c), which show projected views of Ge(111) and CaGe$_2$. For both materials, the Ge layers have the same structure. The difference is that the CaGe$_2$ has an atomic layer of intercalated Ca to separate the individual Ge layers (two flat Ge planes can be considered as an single buckled Ge layer with honeycomb structure). The overall structure of the Ge(111) surface should stay unchanged, as shown by integer order diffraction peaks maintaining their position and intensity throughout the deposition. Thus, the evolution of the RHEED pattern is consistent with the transition from Ge(111) to CaGe$_2$.

To verify that the 1×3 reconstruction is indeed the pattern for $CaGe_2$, we synthesized $CaGe_2$ using the established method of Ca deposition on Ge(111) followed by a post-anneal.[9] Starting with the annealed Ge(111) surface and growing a buffer layer of Ge, we then deposit 20 nm Ca at room temperature. During Ca growth, the RHEED pattern quickly loses its streaks and becomes featureless, indicating an amorphous Ca layer. We then proceed to heat up sample to 750 ˚C to anneal and evaporate extra Ca from the surface. As the sample temperature rises, the amorphous RHEED begins to show integer order diffraction peaks of the Ge(111) lattice around 480 ˚C. As the temperature continues rising, the 1×3 reconstruction peaks appear at around 690 ˚C and remain for the duration of the anneal. The final RHEED pattern for post-annealed $CaGe_2$ is shown in Figure 3. We can see that both directions show the similar 1×3 reconstructed RHEED patterns as our co-deposition growth, so we verify that our co-deposition growth produces $CaGe_2$ films. Another important conclusion is that the RHEED pattern for the co-deposition growth is significantly better than for the post-anneal growth evident from the latter showing slight polycrystalline rings in the pattern. This might be due to the formation of other stable phases such as CaGe in the post-anneal method, which are suppressed in the adsorption-limited growth regime.

Next, we investigate the growth temperature window by performing co-deposition growth at substrate temperatures of 650 °C and 550 °C. Figure 4(a) shows the RHEED patterns for 650 ˚C, while Figure 4(b) shows RHEED patterns for 550 °C. It is clear that as the growth temperature is decreased, the resulting $CaGe_2$ film degrades in quality, as seen from the RHEED pattern losing the 1×3 reconstruction and the integer order streaks starting to exhibit superimposed dots. The dot pattern is indicative of a 3D reciprocal lattice, which emerges as the surface becomes slightly rougher and the RHEED beam diffracts via transmission through islands on the surface. The asymmetry of the dot pattern is due to a slight out-of-plane tilt of the sample holder.

We next focus on the conversion of the $CaGe_2$ films to germanane by submerging in 37% HCl solution (aqueous) at various temperatures (-40 °C to room temperature) and for various times (1 hr to 24 hr).[5, 11] For characterization by x-ray diffraction and Raman spectroscopy, we synthesize a 150 nm $CaGe_2$ film (sample A) and de-intercalate by submerging in HCl at 5 °C for 1 hour. The θ-2θ x-ray diffraction scan for the resulting film shows two peaks (Figure 5b). The peak at $2\theta = 27.3°$ is from the Ge(111) substrate (see Figure 5a), while the peak at $2\theta = 15.2°$ corresponds to the (0001)-oriented germanane.[5] This confirms that the resulting film is germanane. The inset of Figure 5b shows a RHEED pattern taken with beam energy of 30 kV. The diffraction spots form a circular pattern and slide with aziumthal rotation, which indicates a 2D reciprocal lattice. The single RHEED pattern also indicates that the orientation of the germanane film is determined by the underlying Ge(111) substrate. Raman spectroscopy also indicates that the resulting film is germanane. Figure 5c shows the Raman spectra for sample A (orange curve), which is different from the Raman spectrum of bare Ge(111) substrate (black curve) which has a single peak at 297 cm$^{-1}$ Raman shift. For sample A, we observe two large peaks on the Raman spectrum. The peak at 228 cm$^{-1}$ corresponds to the $A_1$ germanane mode while the $E_2$ GeH mode at 302 cm$^{-1}$ can be seen as a kink on the Ge(111) substrate peak. These peaks are consistent with Raman spectra of bulk germanane crystals.[5] Because the $A_1$ peak is specific to germanane (not present in bulk germanium), it is a strong indicator for the presence of germanane. Thus, x-ray diffraction, RHEED, and Raman spectroscopy provide very strong evidence that the resulting film is germanane.

To investigate the morphology of the film, we perform *ex situ* AFM measurements on a thin 5 nm $CaGe_2$ film (sample B) before and after de-intercalating in HCl at -40 °C for 1 hour (Figure 6). For reference, we show an AFM scan of a 5 nm Ge buffer layer on a Ge(111) substrate taken from the same wafer (Figure 6(a)). This Ge buffer layer is grown at 650 °C and then heated to 750 °C to reproduce the conditions just prior to the $CaGe_2$ deposition. AFM of the Ge buffer layer shows the formation of terraces due to the miscut of the Ge(111) substrate. The typical height of each terrace is ~5 nm. Figure 6(b) shows that with the growth of 5 nm $CaGe_2$, the terrace size becomes larger. This indicates the propensity for the $CaGe_2$ to grow flat on the (111) plane. In order to maintain the global miscut of the substrate, the terrace height also increases to typical values of ~10 nm (note that a terrace width of a few microns and height of ~10 nm corresponds to a miscut of ~0.2°). After de-intercalation (Figure 6(c)), the overall terrace size is similar to the $CaGe_2$ film. To determine the local roughness of the film, we analyze the rms roughness within individual terraces. We find that the Ge buffer has rms roughness of 0.26 nm, the $CaGe_2$ film has rms roughness of 0.40 nm, and the germanane film has rms roughness of 0.25 nm. Most importantly, even after submerging in HCl solution to obtain germanane, the individual terraces remain atomically smooth. For reference, we perform RHEED measurements on sample B after HCl processing (Figure 6d). These patterns are similar to sample A (Fig. 5(b) inset) with 2D-like diffraction spots, but are dimmer due to the 15 kV beam voltage. The 2D-like pattern is consistent with the atomic scale smoothness indicated by the AFM measurements.

Auger electron spectroscopy is used to further characterize the films. For reference, Figure 7(a) shows the Auger spectrum of a Ge(111) substrate prepared by chemical etch and UHV anneal to remove surface oxides as discussed earlier. This spectrum is measured *in situ* after the UHV anneal. Important features here are the relatively small oxygen peak at 510 eV and the presence of very weak Ge peaks in the 1000-1200 eV range. There is also a small peak near 260 eV that is likely to come from carbon contamination. Figure 7(b) shows the Auger spectrum taken after co-deposition growth of 10 nm $CaGe_2$ (sample C), giving a clear Ca peak at 294 eV and still maintaining the Ge peaks in the 1100 eV range. This spectrum is taken *ex situ* and the large O peak likely appears due to exposure of $CaGe_2$ films to atmosphere. Next, we perform de-intercalation by HCl at room temperature for 24 hours and subsequently measure the Auger spectrum (Figure 7(c)). The Ca peak disappears from the pattern (within noise limits) while the small peak near 260 eV remains, similar to the fresh Ge(111) surface. Despite the exposure to atmosphere, the O peak is severely reduced, which indicates that the film is relatively resistant to oxidation. For comparison, we perform Auger spectroscopy on sample A after de-intercalation (150 nm thick, 5 °C, 1 hour) and a very similar Auger spectrum is obtained. To compare the de-intercalation at higher temperatures and for longer times, we perform RHEED measurements on sample C and the resulting patterns are shown in Figure 7(d). Unlike sample A (5 °C, 1 hour) and sample B (-40 °C, 1 hour) which exhibit 2D patterns, sample C exhibits 3D-like RHEED patterns (i.e. fixed spots that change in intensity as the sample is rotated in-plane). This indicates that germanane films are smoother for lower temperatures and shorter times of the HCl processing. More detailed studies are needed to better understand the de-intercalation process.

**IV. CONCLUSION**

We have demonstrated the successful co-deposition of CaGe$_2$ films on Ge(111) by MBE. The growth occurs within an adsorption-limited growth regime with overpressure of Ca, which ensures stoichiometry of the film. The best RHEED patterns are obtained at a growth temperature 750 °C. We compare this new growth method with previous methods based on Ca deposition on Ge(111) and post-anneal. We find that the co-deposition method produces RHEED patterns that are free of the polycrystalline rings that occur for the post-anneal growth, which suggests that the co-deposition method inhibits the formation of other stable phases such as CaGe. The CaGe$_2$ films are then immersed in hydrochloric acid and subsequent characterizations by x-ray diffraction and Raman spectroscopy indicate that the resulting film is germanane. RHEED patterns indicate that the crystalline orientation of the germanane film is determined by the underlying Ge(111) substrate. Furthermore, both RHEED and AFM measurements indicate that the local morphology is smooth at the atomic scale. However, the grain size is limited to a few micrometers, primarily due to the miscut of the Ge substrate. Thus, optimization of the substrate could yield much larger grain sizes. These results provide a significant advance toward the long-term goal of developing large area germanane.

**ACKNOWLEDGMENTS**

We acknowledge the technical assistance of Hua Wen and Yunqiu Kelly Luo. This work was supported by ARO (W911NF-11-1-0182), NSF (DMR-MRSEC 0820414), UC Labs (12-LR-239009), ONR (N00014-12-1-0469), and ENCOMM. JEG acknowledges the support of ARO (W911-NF-12-1-0481).

**FIGURE CAPTIONS**

FIG. 1. (a) Atomic structure diagram of germanane (GeH), (b) Projected view of the bulk germanium lattice. The axis indicates the directions within the cubic lattice and in the corresponding hexagonal lattice (for $CaGe_2$ and germanane), (c) Projected view of $CaGe_2$ films, (d) Project view of stacked germanane layers.

FIG. 2. (a) RHEED patterns for a Ge(111) substrate after UHV anneal. Left and right patterns are for the $[11\bar{2}0]$ and $[1\bar{1}00]$ azimuths, respectively, (b) RHEED patterns after the growth of a 5 nm Ge buffer layer, (c) RHEED patterns after the growth of $CaGe_2$ at 750 °C.

FIG. 3. RHEED patterns for $CaGe_2$ prepared by depositing 10 nm Ca onto Ge(111) and post-annealing up to 750 °C in UHV.

FIG. 4. (a) RHEED patterns for co-deposited $CaGe_2$ grown at 650 °C, (b) RHEED patterns for co-deposited $CaGe_2$ grown at 550 °C.

FIG. 5. (a) θ-2θ x-ray diffraction scan of a Ge(111) substrate, (b) θ-2θ x-ray diffraction scan of a 150 nm GeH film (sample A). The peak at 15.2° corresponds to (0001)-oriented GeH. Inset: RHEED pattern of sample A with beam along $[1\bar{1}00]$ and beam energy of 30 kV, (c) Raman spectrum for sample A (orange curve) and Ge(111) substrate (black curve). Data are plotted on a log scale.

FIG. 6. (a) AFM image of Ge(111) substrate, (b) AFM image of 5 nm $CaGe_2$ on Ge(111) (sample B), (c) AFM image of sample B after de-intercalation in HCl solution for 1 hr at -40 °C (i.e. 5 nm GeH), (d) RHEED pattern of 5 nm GeH with beam along $[1\bar{1}00]$.

FIG. 7. (a) *In situ* Auger spectrum for Ge(111) prepared by chemical etch and UHV anneal, (b) *Ex situ* Auger spectrum for a 10 nm CaGe$_2$ film (sample C), (c) Auger spectrum for sample C after HCl immersion at 25 °C for 24 hr, (d) Auger spectrum for sample A after HCl immersion at 5 °C for 1 hr, (e) RHEED patterns for sample C after HCl immersion.

**Figure 1**

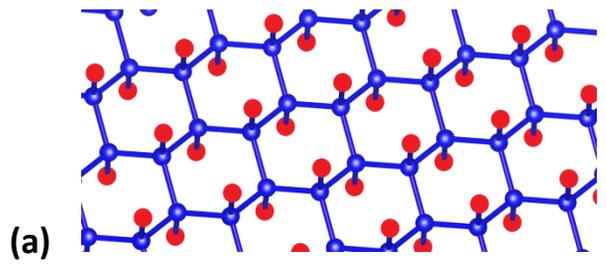

(a)

$[111]_c = [0001]_h$

$[\bar{1}\bar{1}2]_c = [\bar{1}100]_h$

$[\bar{1}10]_c = [11\bar{2}0]_h$

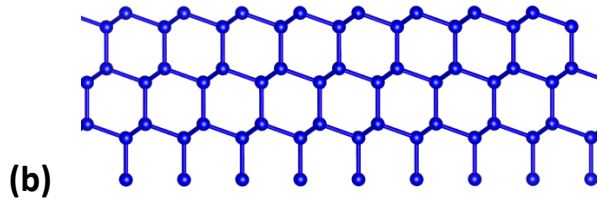

(b)

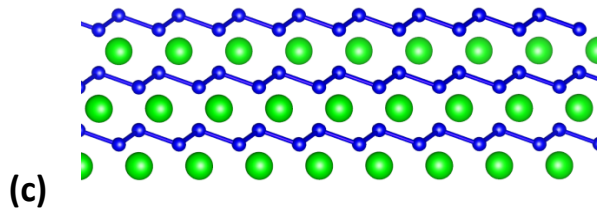

(c)

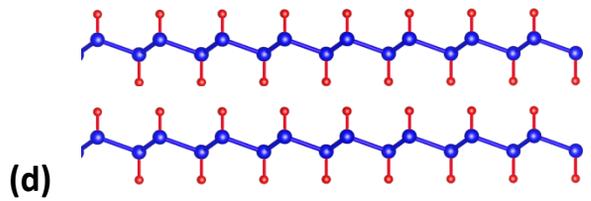

(d)

# Figure 2

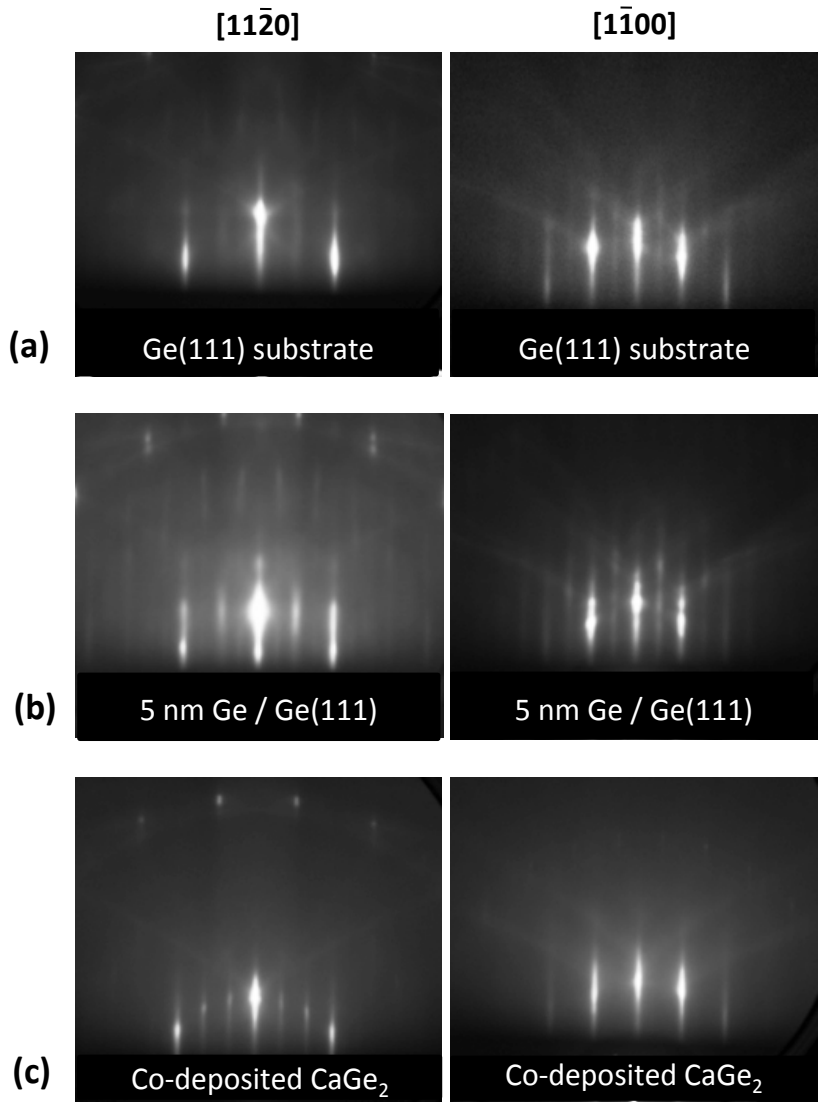

# Figure 3

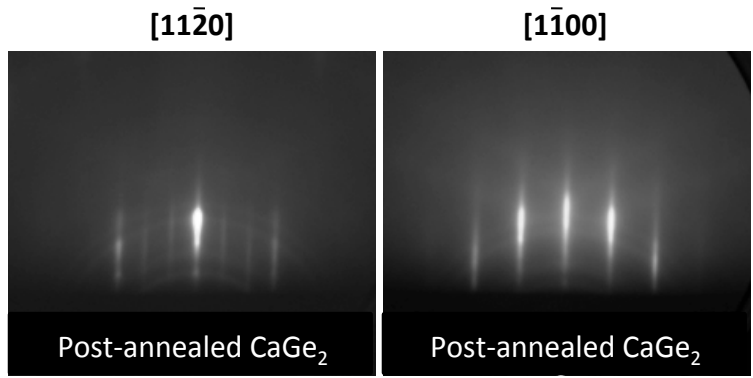

# Figure 4

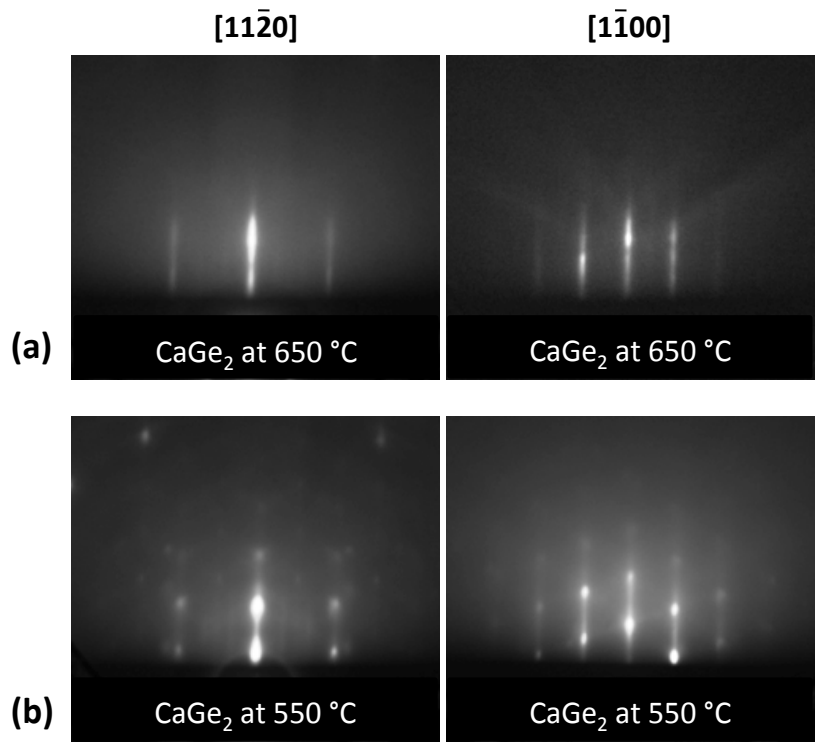

**Figure 5**

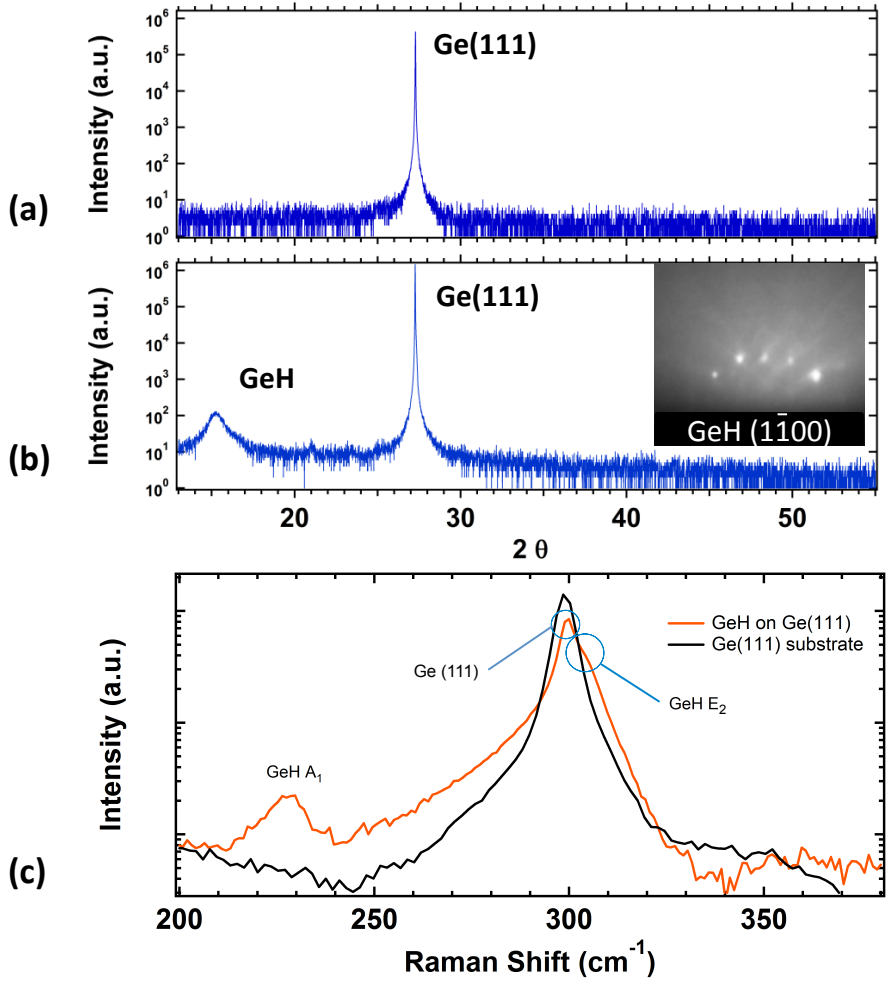

**Figure 6**

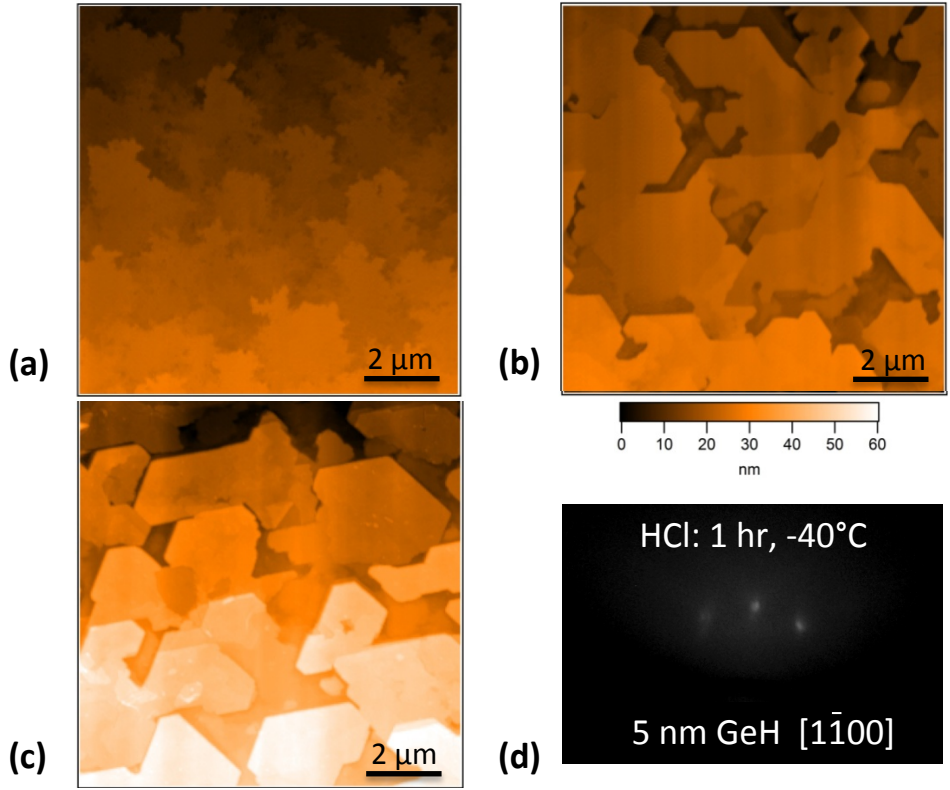

**Figure 7**

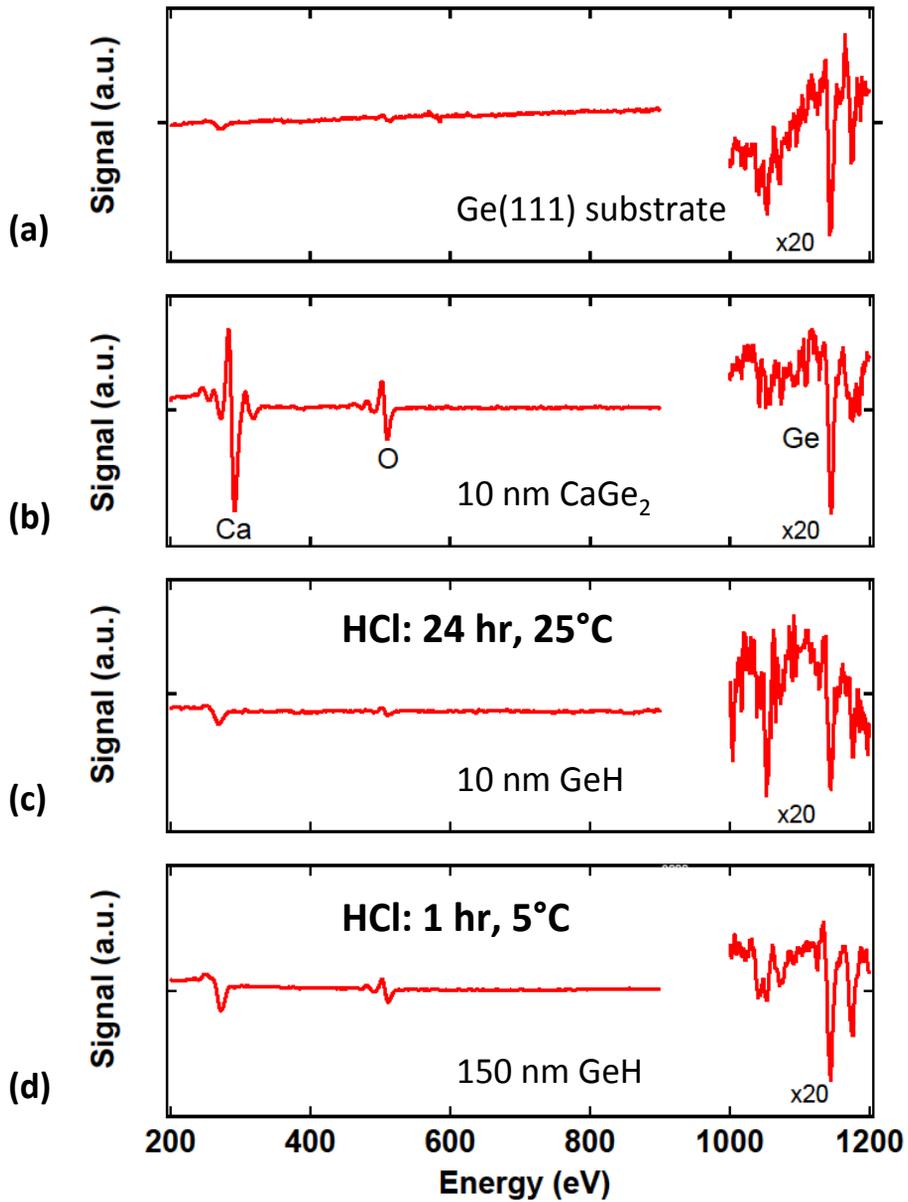
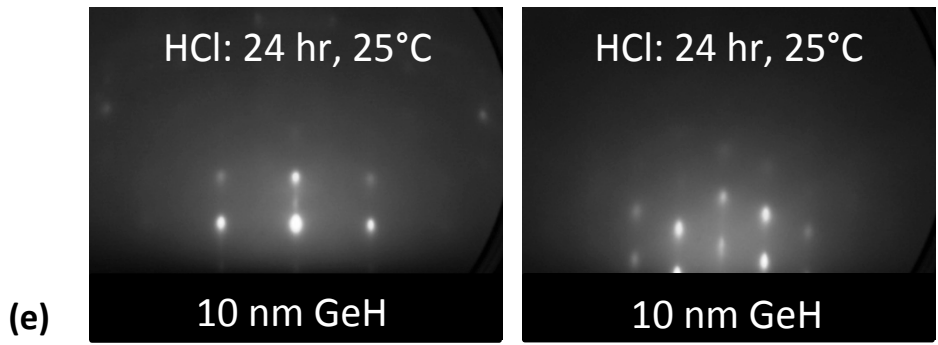